\documentclass[fleqn,usenatbib]{mnras}

\usepackage[T1]{fontenc}
\usepackage{ae,aecompl}

\usepackage{subcaption}
\usepackage{multicol}

\usepackage{graphicx}	
\usepackage{amsmath}	
\usepackage{amssymb}	
\usepackage{natbib}
\usepackage{lscape}

\title[GAN recovery of galaxy features]{Generative Adversarial Networks recover features in astrophysical images of galaxies beyond the deconvolution limit}

\author[K. Schawinski et al.]{
\parbox[t]{18cm}{
Kevin Schawinski,$^{1}$\thanks{E-mail: kevin.schawinski@phys.ethz.ch} Ce Zhang,$^{2}$\thanks{E-mail: ce.zhang@inf.ethz.ch} Hantian Zhang,$^{2}$ Lucas Fowler,$^{1}$ and Gokula Krishnan Santhanam$^{2}$
}\\
\\
$^{1}$Institute for Astronomy, Department of Physics, ETH Zurich, Wolfgang-Pauli-Strasse 27, CH-8093, Z\"{u}rich, Switzerland\\
$^{2}$Systems Group, Department of Computer Science, ETH Zurich, Universit\"{a}tstrasse 6, CH-8006, Z\"{u}rich, Switzerland 
}

\date{Accepted 2017 January 18. Received 2017 January 18; in original form 2016 November 30}

\pubyear{2017}

\begin{document}
\label{firstpage}
\pagerange{\pageref{firstpage}--\pageref{lastpage}}
\maketitle

\begin{abstract}
Observations of astrophysical objects such as galaxies are limited by various sources of random and systematic noise from the sky background, the optical system of the telescope and the detector used to record the data. Conventional deconvolution techniques are limited in their ability to recover features in imaging data by the Shannon-Nyquist sampling theorem. Here we train a generative adversarial network (GAN) on a sample of $4,550$ images of nearby galaxies at $0.01<z<0.02$ from the Sloan Digital Sky Survey and conduct $10\times$ cross validation to evaluate the results. We present a method using a GAN trained on galaxy images that can recover features from artificially degraded images with worse seeing and higher noise than the original with a performance which far exceeds simple deconvolution. The ability to better recover detailed features such as galaxy morphology from low-signal-to-noise and low angular resolution imaging data significantly increases our ability to study existing data sets of astrophysical objects as well as future observations with observatories such as the Large Synoptic Sky Telescope (LSST) and the \textit{Hubble} and \textit{James Webb} space telescopes.
\end{abstract}




\section{Introduction}
Any telescope observation of astrophysical objects is limited by the noise in the image, driven both by the detector used and the sky background. Similarly, the observation is limited in angular resolution by the resolving power of the telescope ($R\sim \lambda/D$) and, if taken from the ground, by the distortions caused by the moving atmosphere (the ``seeing''). The total blurring introduced by the combination of the telescope and the atmosphere is described by the point spread function (PSF). An image taken by a telescope can therefore be thought of as a convolution of the true light distribution with this point spread function plus the addition of various sources of noise. The Shannon-Nyquist sampling theorem \citep{nyquist1928certain, shannon49} limits the ability of deconvolution techniques in removing the effect of the PSF, particularly in the presence of noise \citep{1998ApJ...494..472M, 1999PhDT........18C, 2002PASP..114.1051S}. 

Deconvolution has long been known as an ``ill-posed'' inverse problem  because there is often no unique solution if one follows the signal processing approach of backwards modelling \citep{1998ApJ...494..472M, 2007A&A...461..373M, 2007MNRAS.378...83L, 2008MmSAI..79.1251L, 2016A&A...589A..81C}. Another standard  practice in tackling inverse problems like these is integrating priors using domain knowledge in forward modelling. For example, if the algorithm knows what a galaxy should look like or it knows the output needs to have certain properties such as being ``sharp'', it will make more informative decisions when choosing among all possible solutions. In this paper we demonstrate a method using machine learning to automatically introduce such priors. This method can reliably recover features in images of galaxies. We find that machine learning techniques can go beyond this limitation of deconvolutions --- by training on higher quality data, a machine learning system can learn to recover information from poor quality data by effectively building priors. 

\section{Method}
Our general method is agnostic as to the specific machine learning algorithm used. In this paper, we choose to use conditional Generative Adversarial Networks (GAN), a state-of-the-art deep learning algorithm for image-to-image translation.
In this work, we adopted a standard GAN architecture; therefore we
only briefly introduce GAN and interested readers can consult \cite{Reed:2016:ArXiv} and \cite{ Goodfellow:2014:NIPS} for details.

In the training phase, the GAN takes as input a set of image pairs---in our case, one image which is degraded (by this we mean: convolved with a worse PSF, or blurred, and with added noise) and the same image without such degradation. The GAN then tries to ``learn'' to recover the degraded image by minimizing the difference between the recovered image and the non-degraded image. 
The function that measures the difference between the two images, which is often called the loss function, is often something simple such as the Euclid distance but can be a more sophisticated function. In the case of a GAN, this function is another neural network (hence the name adversarial) whose goal is to distinguish the recovered image from a non-degraded image. These two neural networks are trained at the same time. This allows the system to learn sophisticated loss functions automatically without hand-engineering.\footnote{It is known that if one uses Euclid distance for image recovery, this often produces blurred images because Euclid distance is uniform over the whole image~\citep{Reed:2016:ArXiv}, thus a more sophisticated loss function could improve the system.} In the testing phase, the GAN takes a different set of degraded images and recovers them.

\begin{figure}
\centering
\includegraphics[width=1.0\linewidth]{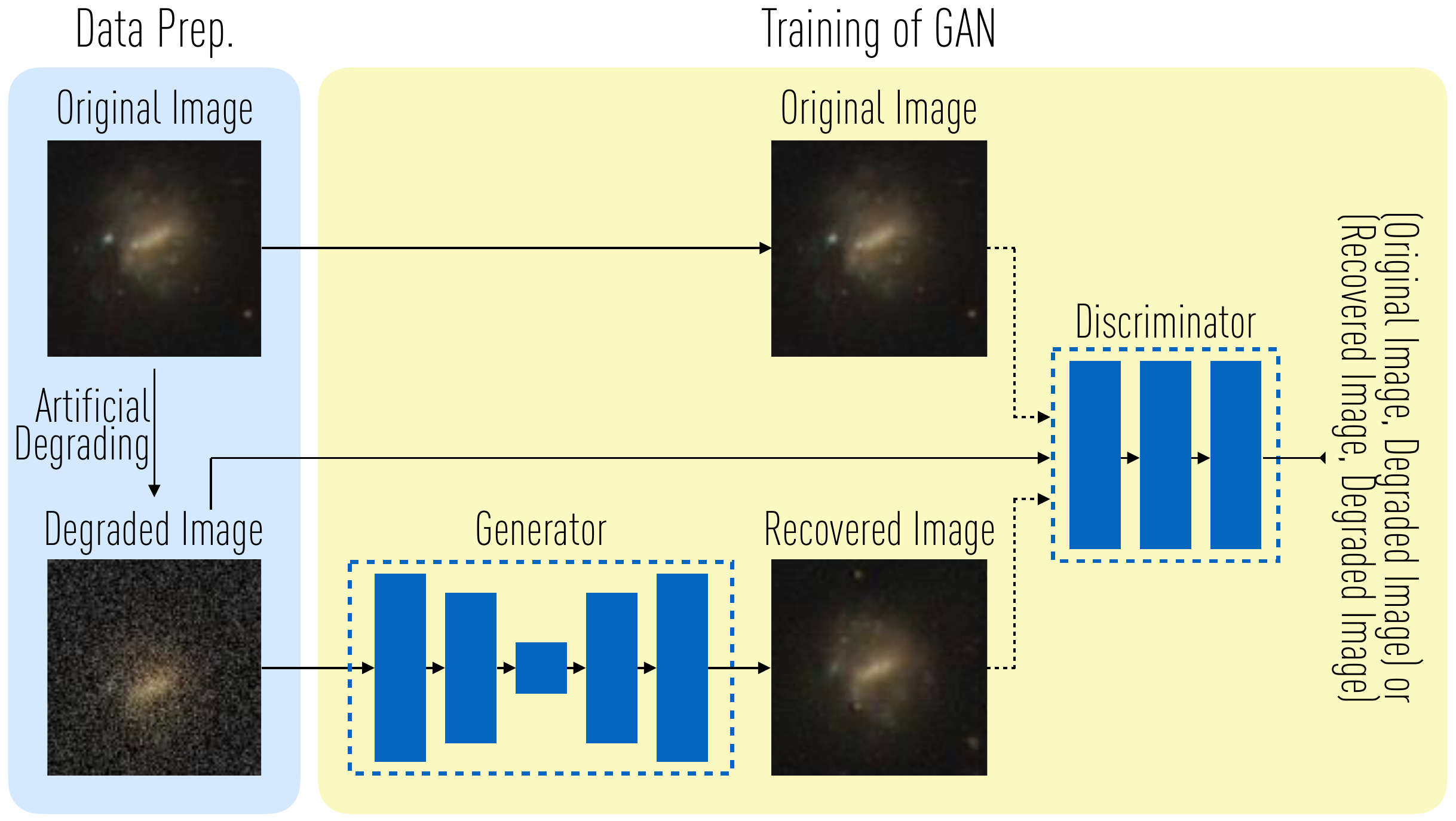}
\caption{Schematic illustration of the training process of our method. The input is a set of original images. From these we automatically generate degraded images, and train a Generative Adversarial Network. In the testing phase, only the generator will be used to recover images.}
\label{fig:gan}
\end{figure}

One remaining challenge is how to generate pairs of images with and without degradation for the training phase. In our framework, we take advantage of the centuries of study of the noise introduced by the telescope and the atmosphere to  {\em weakly supervise} a GAN network by {\em simulating} the blurring process automatically. This allows us to easily harvest a large training set automatically without any human intervention. Furthermore, it allows us to automatically scale our system, and arguably achieve better quality, when future large-scale sky survey data from e.g. LSST  \citep{2009arXiv0912.0201L} or \textit{Euclid} \citep{2011arXiv1110.3193L} are available. We outline the method in Figure \ref{fig:gan}.

\begin{figure*}
\centering
\includegraphics[width=\linewidth]{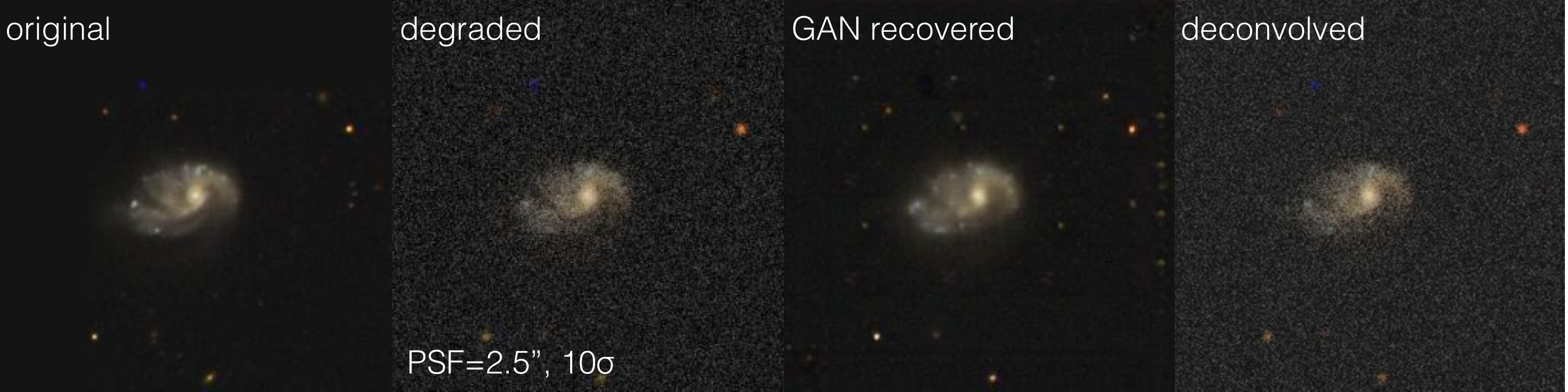}

\caption{We show the results obtained for one example galaxy. From left to right: the original SDSS image, the degraded image with a worse PSF and higher noise level (indicating the PSF and noise level used), the image as recovered by the GAN, and for comparison, the result of a deconvolution. This figure visually illustrates the GAN's ability to recover features which conventional deconvolutions cannot.}
\label{fig:main_example}
\end{figure*}

We select a sample of 4,550 galaxies from the Sloan Digital Sky Survey Data Release 12 \citep{2000AJ....120.1579Y, 2015ApJS..219...12A} in the redshift range $0.01<z<0.02$ and conduct 10$\times$ cross validation for all of our experiments (each fold contains 4,105 images for training and 455 for testing). We obtain the $g$, $r$ and $i$-band images for these objects and process them using an asinh stretch ($y=\rm{asinh}(10x)/3$) to produce 3-band RGB images. The transform to asinh stretch rather than keeping the linear data follows the best practice of making the range of input values for a neural network comparable across images. This step has been shown to help make the training procedure faster and easier in other applications \citep{Sola:1997:NuclearScience}. We note that this process involved clipping extreme pixel values and in principle makes it impossible to fully recover the original flux calibration; exploring how to avoid this while not degrading the neural net training performance is an interesting project. 

In order to test the performance of the GAN, we generate a grid of training sets from the galaxy sample. In each training set, we convolve the images with a Gaussian PSF with a full-width at half-maximum (FWHM) of FWHM=$\left[1.4, 1.8, 2.0, 2.5\right]\arcsec $. The median seeing of SDSS images is $\sim1.4\arcsec$ so we explore images of effectively the same resolution all the way to a significantly worse seeing of $2.5\arcsec$. After convolving the images with a Gaussian filter representing worse seeing, we adjust the noise level, first restoring it to that of the original image, and then increasing it so that $\sigma_{\rm new}=[1.0, 1.2, 2.0, 5.0, 10.0]\sigma_{\rm original}$ to mimic shallower images. We train the GAN using open source code released by \cite{Reed:2016:ArXiv} with TITAN X PASCAL GPUs. Training finishes in 2 hours per setting per fold (200 hours in total).

\begin{figure*}
\centering
\includegraphics[width=0.49\linewidth]{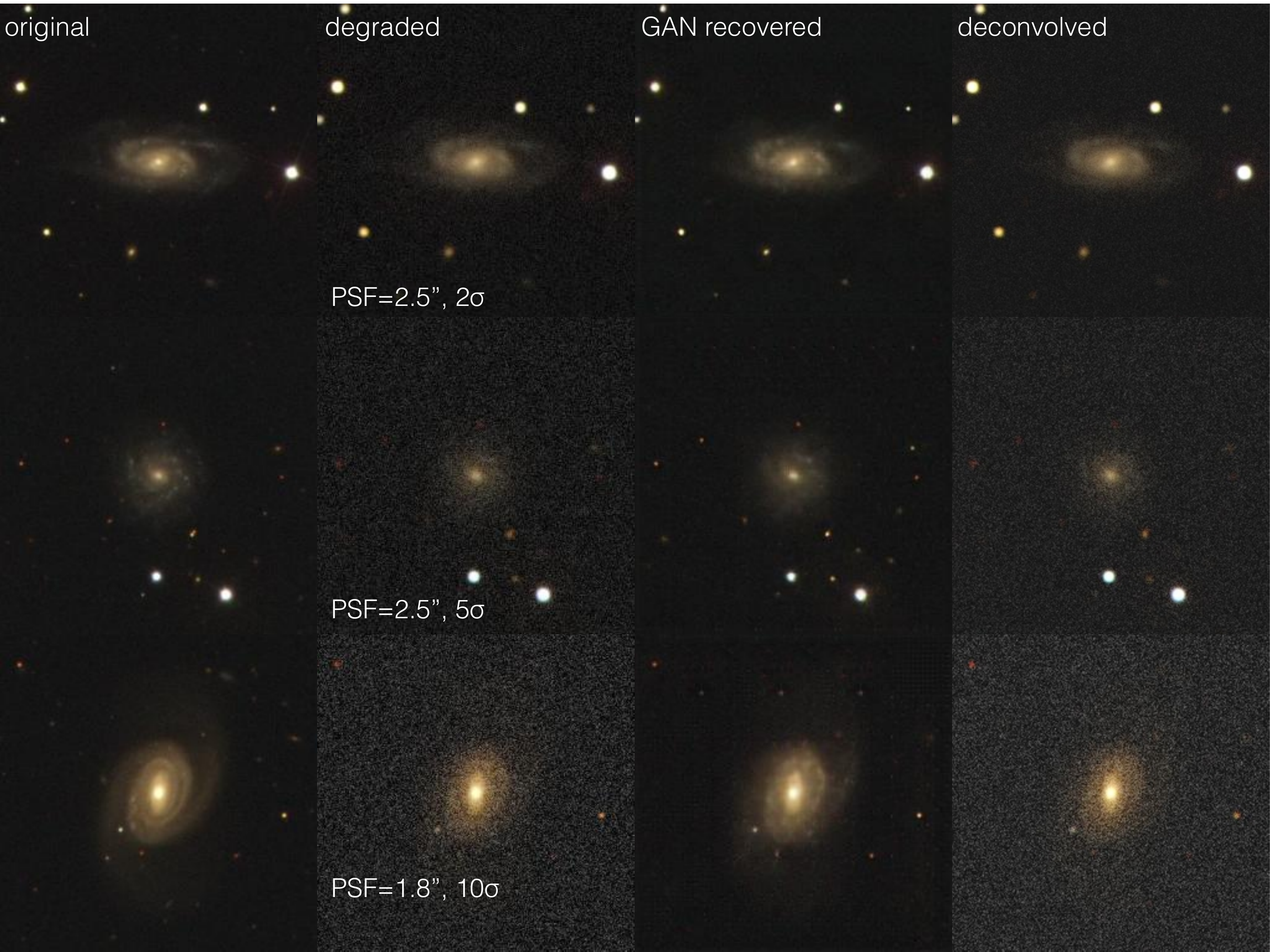}
\includegraphics[width=0.49\linewidth]{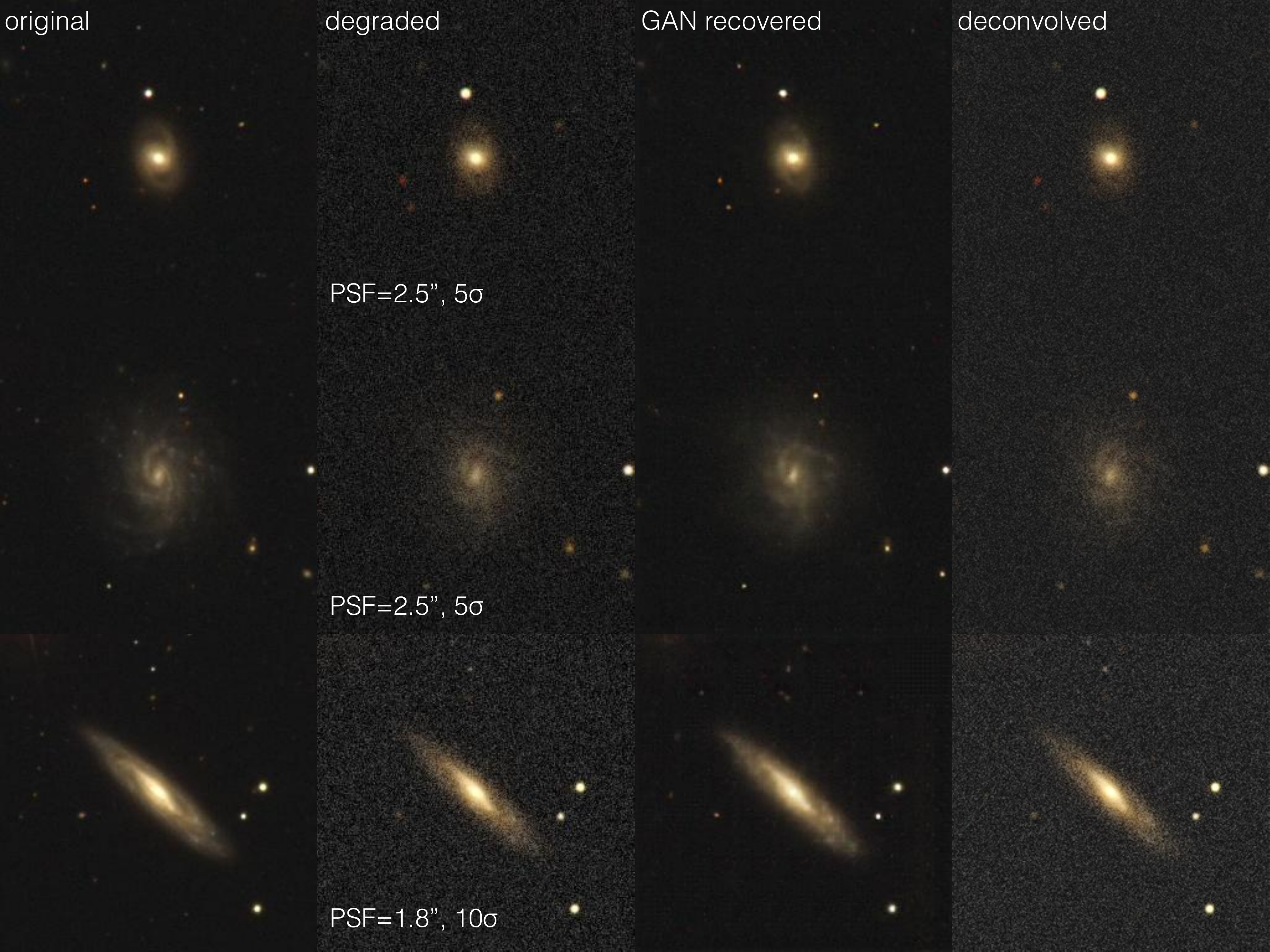}
\includegraphics[width=0.49\linewidth]{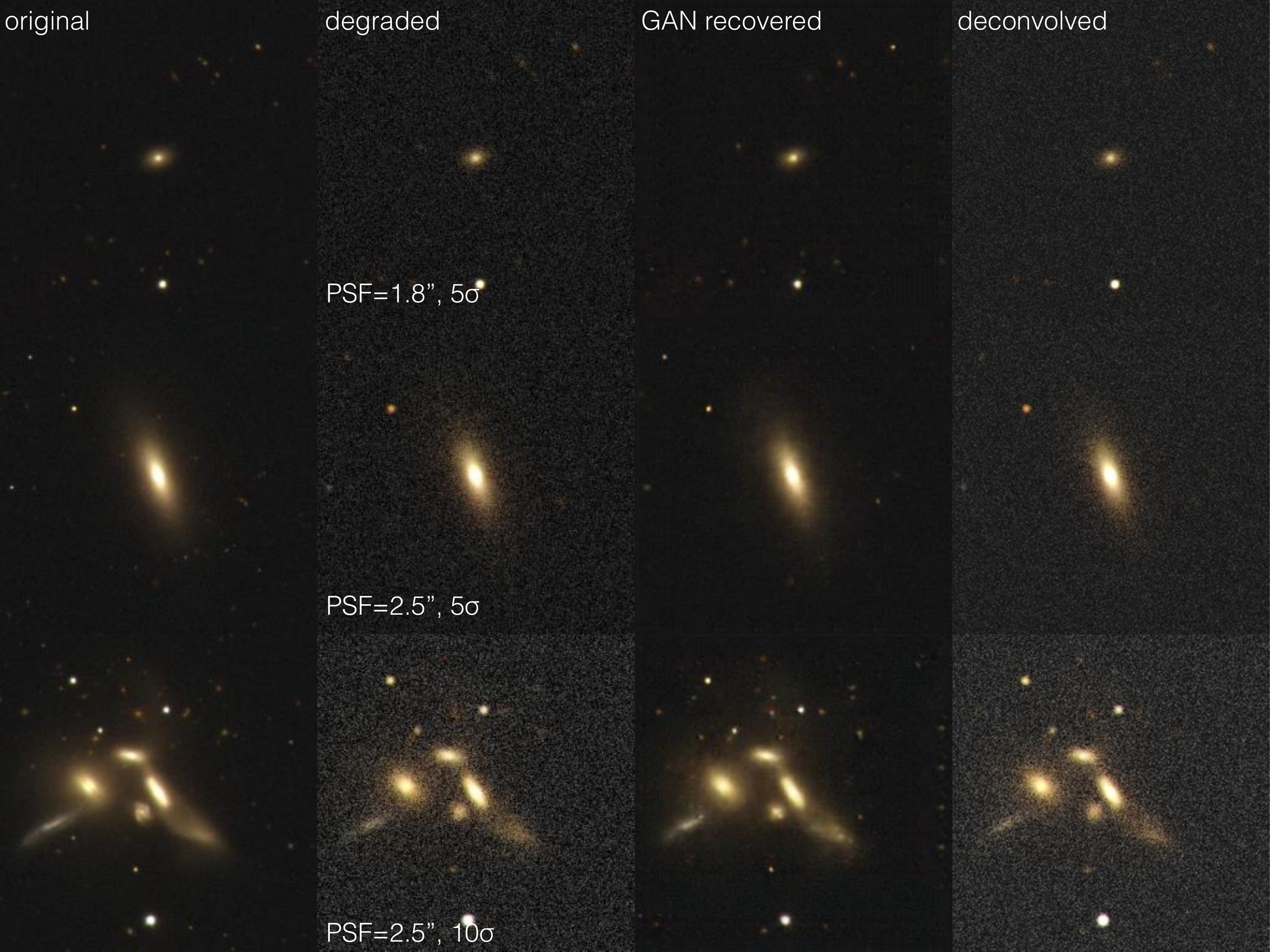}
\includegraphics[width=0.49\linewidth]{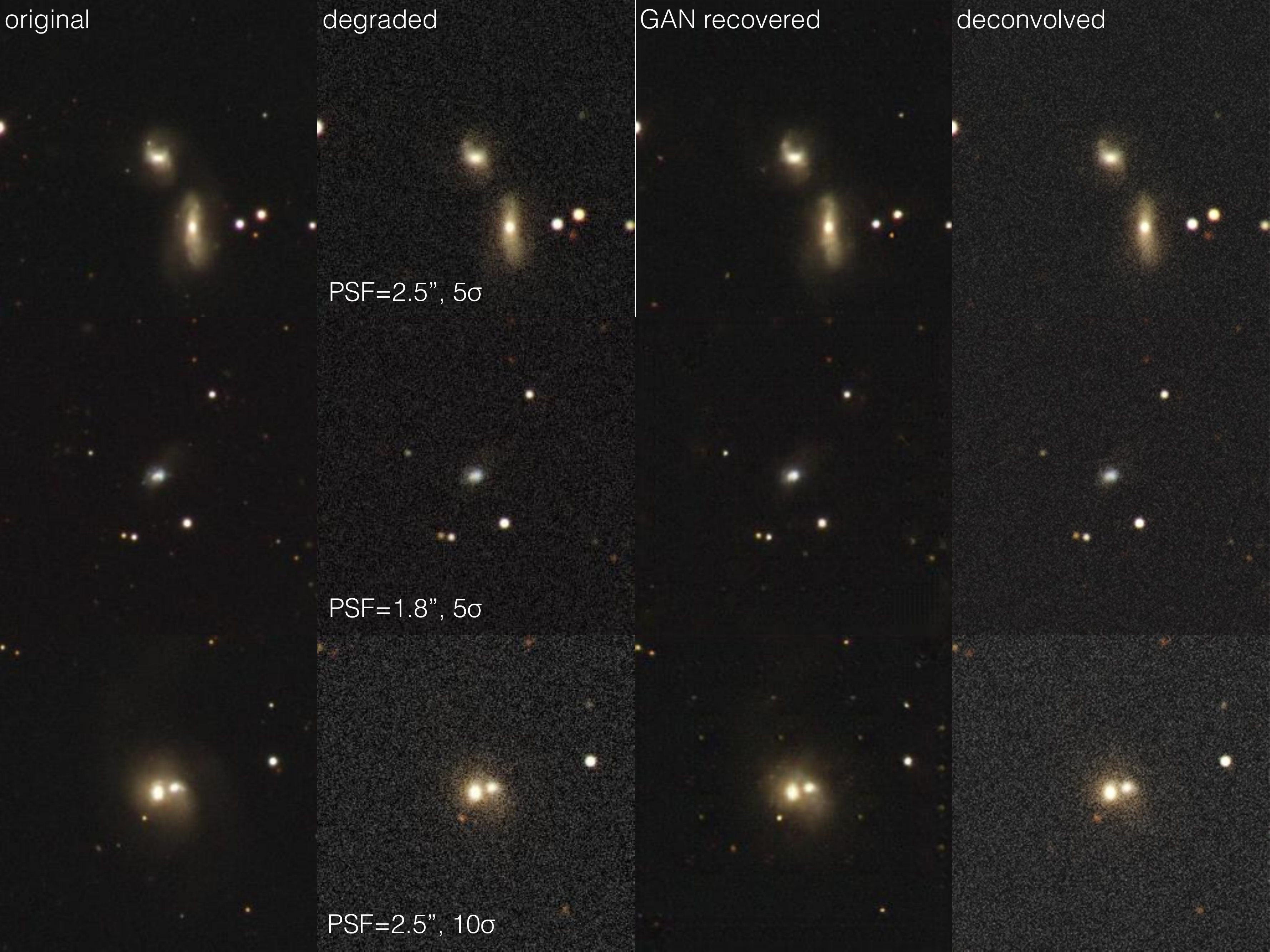}

\caption{We show some further representative results for different galaxy types and with various levels of degradation. In each row, we show with the same layout as Figure \ref{fig:main_example}. Since the GAN has been trained on images of galaxies with similar properties, it is able to recover details which the deconvolution cannot, such as star-forming regions, dust lanes and the shape of spiral arms. The top two panels are examples of spiral galaxies. The bottom-left panel shows early-type galaxies (including a dense cluster). The bottom-right panel shows galaxy mergers; note in particular that the GAN reconstruction makes it easier to identify these systems as merging, as opposed to being undisturbed or superpositions . For more detailed results, see Appendix \ref{sec:gan_ouput}.}
\label{fig:results}
\end{figure*}

\begin{figure}
\centering
\includegraphics[width=\linewidth]{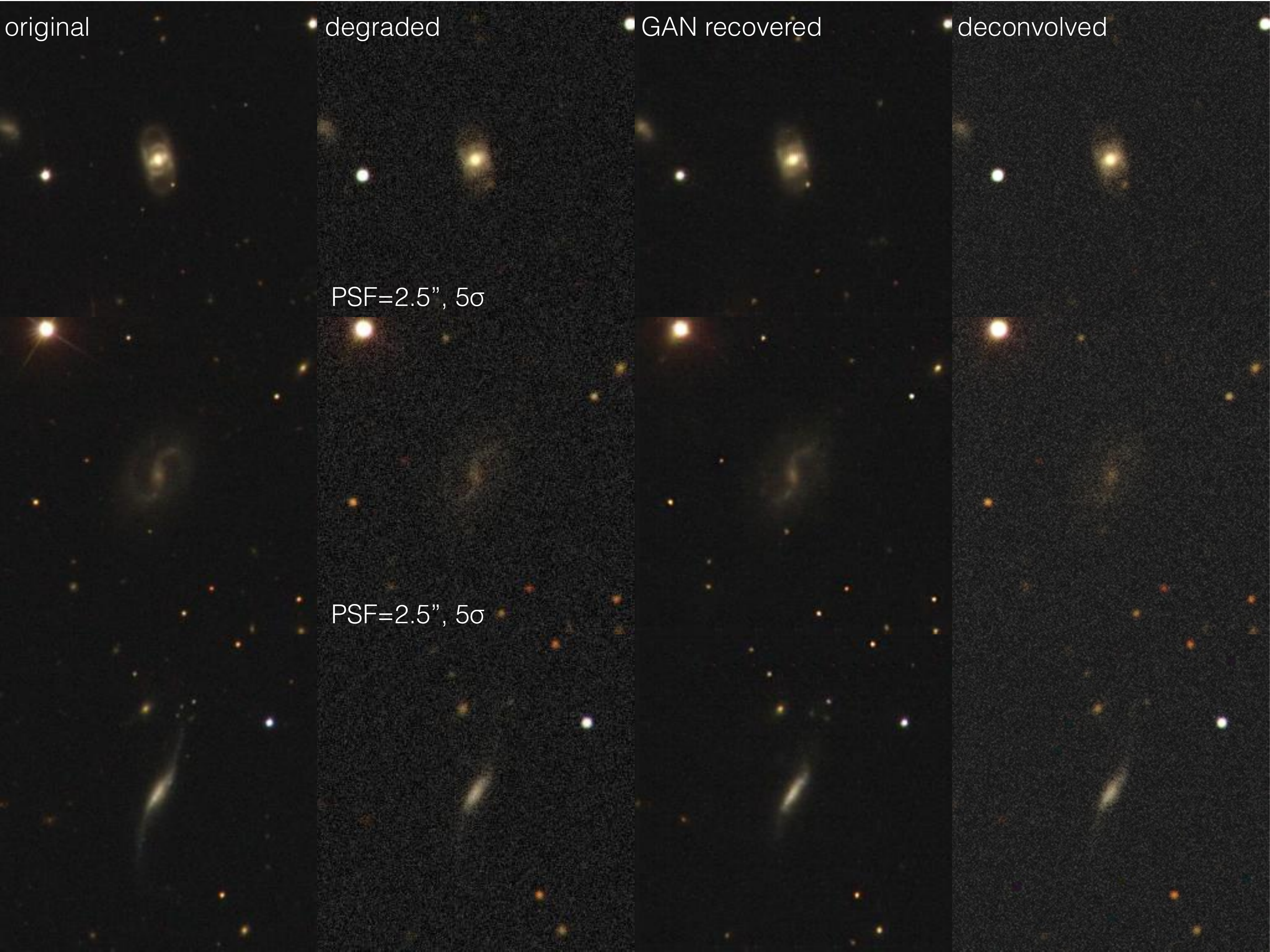}

\caption{We show three examples where the GAN reconstruction fails at some level. Two of the three can be accounted for by the fact that the objects in question are rare, and so the training set did not sufficiently prepare the GAN to deal with them. In the top row is a rare kinematic structure \citep{1996FCPh...17...95B}. The middle row is a barred spiral whose outer arms were so far below the noise that they could not be reconstructed. The bottom row is a tidally warped edge-on disk, a class of galaxies sufficiently rare that the training did not prepare the GAN to recognise it. }
\label{fig:failures}
\end{figure}

{
\begin{table}
\centering

\begin{subtable}{.9\linewidth}
\centering
\begin{tabular}{r|rrrr}
\hline
White noise level  & \multicolumn{4}{c}{Gaussian PSF FWHM }\\
relative to        & \multicolumn{4}{c}{parameter in arcsec}\\
original image     & 1.4$\arcsec$ & 1.8$\arcsec$ & 2.2$\arcsec$ & 2.5$\arcsec$ \\
\hline
1.0$\sigma_{\rm original}$    & 39.6    & 39.8   & 39.2   &  38.9  \\
1.2$\sigma_{\rm original}$    & 40.0    & 39.5   & 39.5   & 38.9   \\
2.0$\sigma_{\rm original}$    & 38.9    & 39.4   & 39.2   & 38.8   \\
5.0$\sigma_{\rm original}$    & 37.8    & 38.1   & 38.1  & 37.8   \\
10.0$\sigma_{\rm original}$   & 35.4    & 37.3   & 37.3   & 36.9   \\
\hline
\end{tabular}
\caption{PSNR (dB) of GAN}
\end{subtable}

\begin{subtable}{.9\linewidth}
\centering
\begin{tabular}{r|rrrr}
\hline
White noise level  & \multicolumn{4}{c}{Gaussian PSF FWHM }\\
relative to        & \multicolumn{4}{c}{parameter in arcsec}\\
original image     & 1.4$\arcsec$ & 1.8$\arcsec$ & 2.2$\arcsec$ & 2.5$\arcsec$ \\
\hline
1.0$\sigma_{\rm original}$   & 36.2    & 35.7   & 35.0  & 34.5  \\
1.2$\sigma_{\rm original}$   & 35.1    & 34.7   & 34.0   & 33.6   \\
2.0$\sigma_{\rm original}$   & 31.6    & 31.4   & 31.1   & 30.9   \\
5.0$\sigma_{\rm original}$   & 24.9    & 24.8   & 24.8   & 25.4   \\
10.0$\sigma_{\rm original}$  & 20.0    & 20.0   & 20.0   & 20.6   \\
\hline
\end{tabular}
\caption{PSNR (dB) of Blind Deconvolution}
\end{subtable}

\begin{subtable}{.9\linewidth}
\centering
\begin{tabular}{r|rrrr}
\hline
White noise level  & \multicolumn{4}{c}{Gaussian PSF FWHM }\\
relative to        & \multicolumn{4}{c}{parameter in arcsec}\\
original image     & 1.4$\arcsec$ & 1.8$\arcsec$ & 2.2$\arcsec$ & 2.5$\arcsec$ \\
\hline
1.0$\sigma_{\rm original}$   & 34.7    & 34.8   & 34.9  & 35.0  \\
1.2$\sigma_{\rm original}$   & 33.3    & 33.5   & 33.6   & 33.7   \\
2.0$\sigma_{\rm original}$   & 29.5    & 29.7   & 29.9   & 30.0   \\
5.0$\sigma_{\rm original}$   & 23.0    & 23.1   & 23.3   & 23.5   \\
10.0$\sigma_{\rm original}$  & 18.8    & 18.8   & 18.9   & 19.0   \\
\hline
\end{tabular}
\caption{PSNR (dB) of Lucy-Richardson Deconvolution}
\end{subtable}

\begin{subtable}{.9\linewidth}
\centering
\begin{tabular}{r|rrrr}
\hline
\# Images & 5 & 10 & 100 & 2000\\
\hline
PSNR & 34.6 & 36.0 & 37.7 & 37.9 \\
\hline
\end{tabular}
\caption{Impact of the Size of Training Set}
\end{subtable}

\caption{(a)-(c) PSNR of images recovered by GAN, Blind Deconvolution, and Lucy-Richardson Deconvolution. (d) Impact of the size of training set on the GAN.}
\label{tab:psnr}
\end{table}
}


\section{Results}

We evaluate our method both quantitatively and qualitatively. Quantitatively, we measure the Peak Signal Noise Ratio (PSNR, \citealt{Xu:2014:NIPS}) of the blurred image and the recovered image.  The PSNR is a popular quantitative measure for image recovery~\citep{Xu:2014:NIPS}. It is defined as the ratio between the maximum possible power of a signal (original image) and the power of the noise (difference between recovered and original image). When the blurred image contains much noise, a case that is known to be a challenge for deconvolution-based approaches, our method can achieve a PSNR of 37.2dB. Blind deconvolution~\citep{bell1995information} \footnote{We compare with blind deconvolution in Matlab. \url{https://www.mathworks.com/help/images/ref/deconvblind.html}} achieves a PSNR of 19.9dB and Lucy-Richardson deconvolution~\citep{richardson1972bayesian,lucy1974iterative} \footnote{We compare with Lucy-Richardson deconvolution in Matlab. \url{https://www.mathworks.com/help/images/ref/deconvlucy.html}} achieves a PSNR of 18.7dB. We compare
our and classic deconvolution approaches in Table \ref{tab:psnr}. We also perform an experiment on the impact of the training set size, reported in Table \ref{tab:psnr}d: the PSNR achieved increases as we increase the training set size from 5 images to 2,000.

For the qualitative analysis, we show example results in Figure \ref{fig:main_example} and \ref{fig:results} where we show the original image, the degraded image with additional noise and a larger PSF, the recovered image and the deconvolved image. We show sample spiral galaxies, early-type galaxies and galaxy mergers, each selected with various levels of degradation. These sample images show what our method is able to recover, and contrast it to the performance of deconvolution. 

In Figure \ref{fig:failures}, we show some examples where the reconstruction method had problems. We stress that these are rare, and we present the results for all parts of the degradation grid evaluated on test galaxies in Appendix \ref{sec:gan_ouput}, which contains the full output as online-only In general, the GAN fails on rare objects which were absent or low in number in the training set, stressing that the performance of our method is strongly tied to the training set; it cannot reconstruct features it has not learned to recognize.

\section{Discussion}
Our method and those like it naturally suffer from some limitations. The main limitation is that the training set ultimately limits the ability to recover features. We trained on galaxies in the nearby universe, and applied the resulting training to similar galaxies in the same redshift range taken with the same camera under similar conditions. If one were to apply it to galaxies at higher redshift, one could simulate effects such as k-correction and cosmological dimming, but the intrinsic morphology of galaxies at say $z\sim2$ or $z\sim6$ are fundamentally different from those at $z\sim0$. One solution to this issue would be to train on images from simulations of galaxy formation at the appropriate epoch. Another limitation of our method is that sophisticated details such as weak lensing shear are impossible to recover via this route, as subtle distortions are truly irrecoverable. Similarly, if there are rare objects absent from the training set, the method may fail, as illustrated in Figure \ref{fig:failures}.

We have shown that it is possible to recover features from low quality imaging data of galaxies which are not possible to recover with a deconvolution. Despite the limitations, our method has enormous scope for application in astrophysics research as it increases the potential for research and discovery. Provided that a suitable training set is available, our method and further developments based on it can be applied to a wide range of existing and future data from SDSS, \textit{Euclid} \citep{2011arXiv1110.3193L} and the LSST \citep{2009arXiv0912.0201L}. For example, using training sets from simulations, one could push to high redshift observation with the \textit{Hubble} and \textit{James Webb} space telescopes to analyze galaxies in the early universe. 

We make all the code available and provide instructions to access a virtual machine which can reproduce the entire analysis at \texttt{http://space.ml/proj/GalaxyGAN.html}.

\section*{Acknowledgements}

KS acknowledges support from Swiss National Science Foundation Grants PP00P2\_138979 and PP00P2\_166159 and the ETH Zurich Department of Physics.  CZ gratefully acknowledge the support from NVIDIA Corporation for its GPU donation, Microsoft Azure for Research award program, and the ETH Zurich Department of Computer Science. We thank William Keel for comments and the anonymous referee for helpful feedback.



\bibliographystyle{mnras}



\appendix

\vspace{-0.5cm}
\section{Full GAN test output}
\label{sec:gan_ouput}
\begin{figure*}
\centering
\includegraphics[angle=90,width=0.8\linewidth]{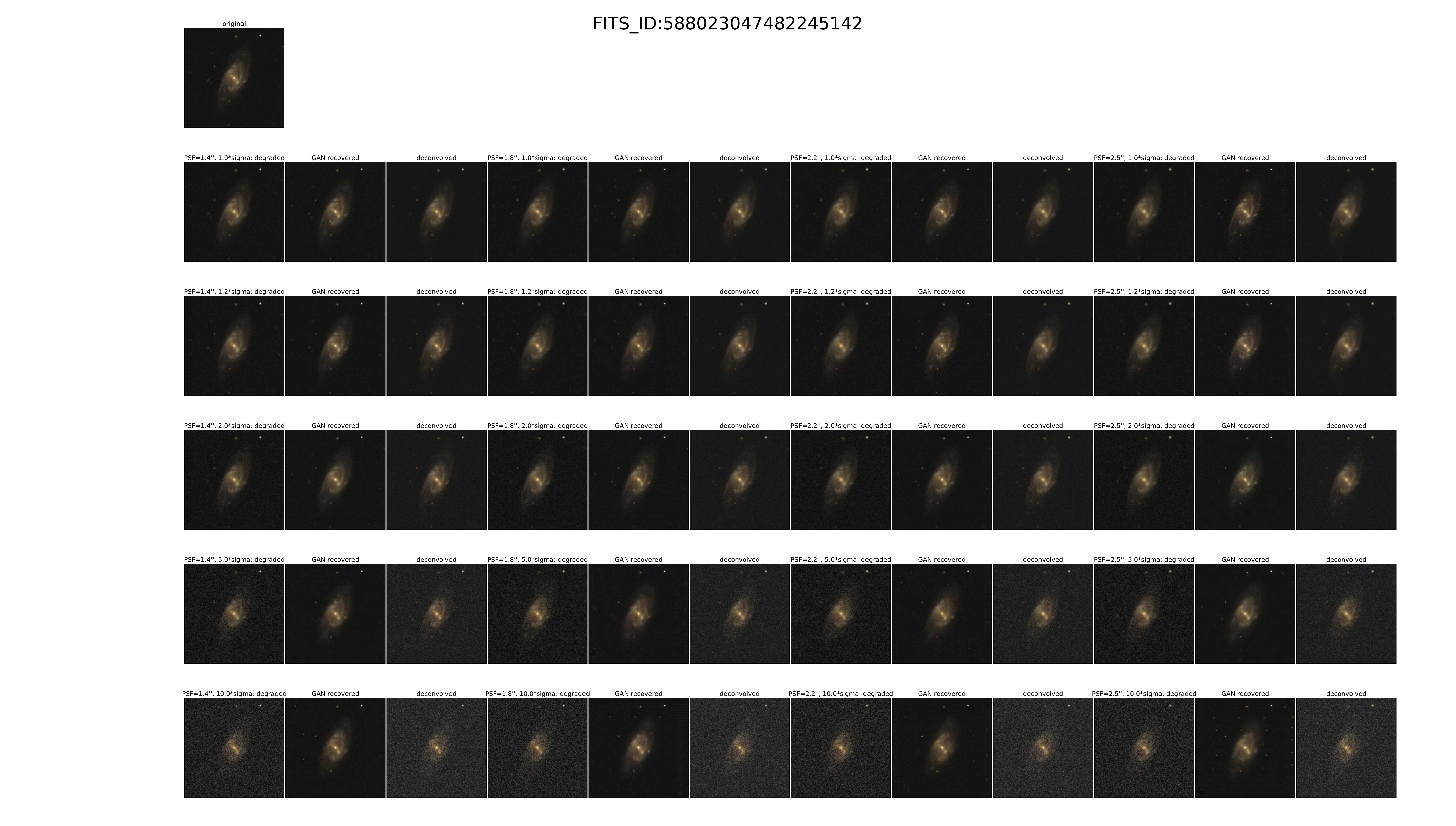}
\vspace{-3.0cm}
\caption{Full outputs for the entire test grid of PSF FWHM and noise levels, for one object. We only show the first object, the remaining outputs are included as online-only. The full set, including the 10x cross validation, is available at \texttt{http://space.ml/supp/GalaxyGAN.html} }
\label{fig:appendix}
\end{figure*}

We present the full evaluation results for all parameter sets in the space of FWHM=$\left[1.4, 1.8, 2.0, 2.5\arcsec \right]$ and $\sigma_{\rm new}=[1.0, 1.2, 2.0, 5.0, 10.0]\sigma_{\rm original}$ in Figure \ref{fig:appendix}. We show an example in Figure \ref{fig:results} and provide the full set, including the 10x cross validation, is available at \texttt{http://space.ml/supp/GalaxyGAN.html}.

\bsp	
\label{lastpage}
\end{document}